# Study and Development of a New Symmetric Key Management Scheme for Wireless Sensor Networks

Yassine Maleh
Dept. Mathematics and Informatics, LAVETE Laboratory
Hassan 1st University
Settat, Morocco
Y_maleh@uhp.org.ma

Abdellah Ezzati
Dept. Mathematics and Informatics, LAVETE Laboratory
Hassan 1st University
Settat, Morocco
abdezzati@uhp.ac.ma

*Abstract*— **Wireless Sensor Network (WSN) is consisting of independent and distributed sensors to monitor physical or environmental conditions, such as temperature, sound, pressure, etc. However, the limited resources of sensors and hostile environments in which they could be deployed, make this type of networks vulnerable to several types of attacks similar to those occurring in ad hoc networks. The most crucial and fundamental challenge that WSN is facing is security. The primary subject of all my work is to address this issue. Due to minimum capacity in-term of memory cost, processing and physical accessibility to sensors devices the security attacks are problematic. They are mostly deployed in open area so are more exposed to different kinds of attacks. They must be designed in a way to successfully recover itself from different kinds of attacks. In this paper, we proposed a new lightweight cryptography algorithm based on LEAP+. Our evaluations on TOSSIM give a precise and detailed idea of the extra cost of consumption of resources needed to ensure the high level of expected security.**

## I. INTRODUCTION

Inexpensiveness, energy efficiency, and consistency in performance is the need of the day for electronic communication. This has led to the advancement of wireless technologies and micro-electro-mechanical systems (MEMS). Which in turn made it possible to make low power tiny devices that runs autonomously. Which evolve a new class of distributed networking named wireless sensor networks (WSNs) [1] [2].

Today we find this kind of network in a wide range of potential applications, including security and surveillance, control, actuation and maintenance of complex systems and fine-grain monitoring of indoor and outdoor environments. The majority of these applications are deployed to monitor an area and to have a reaction when they register a critical event. The data does not need to be confidential in areas such as capturing indoor and outdoor environmental events. However, the confidentiality of data can be essential in other applications, Security is the most important issue for WSN because of the fact of its application. It can be setup in a very critical systems for example in hospitals, airports, military applications, burglar alarms, environment control, smart homes and traffic surveillance. The classic security mechanism has to be avoided in WSN due to limited constraints. It is therefore challenging to come up with a new technique that address the issue while keeping in mind the limitations of WSN that we have discussed in the introduction chapter. It's really important for security to know which specific thing should be protected and in in which way to protect. Sensor networks are vulnerable against external and internal attacks due to their unique characteristics like

- They can be physically access, they are mostly deployed in an open area so any machine or human user can easily reprogram or destroy them.
- The sensor nodes are limited with computation, communication, memory and power capabilities which make it difficult to use cryptographic algorithms and protocols for the required security services.
- As the communication channels is public so any external or internal device can have access to the information exchange.
- The monitoring and control of the actual states of the elements of the network is difficult due to the inherent distributed nature of sensor networks. A failure or the cause of failure in any elements may remain hidden. In order to achieve this in the next section chapter we will present cryptographic primitives and models for WSN in literature.

The paper is structured as following. In the next section, we present in related work. Section 3 discus security model for WSNs. Section 4 describes proposed model to secure the global key. In the Section 5, we present performances analyses and discussion of our scheme. Finally, the paper ends with a conclusion and future works.

## II. RELATED WORKS

At present, there are many security schemes designed for general ad hoc network, but very few for wireless sensor network specifically. In this section, we describe three typical security protocols for sensor network and summarize their features.

Perig et al [3] proposed SPINS protocol which is a suite of security building blocks proposed. It is optimized for resource constrained environments and wireless communication. SPINS has two secure building blocks: SNEP and µTESLA. SNEP uses a shared counter between the two communicating parties and applies the counter in calculating encryption and a message authentication code (MAC) to provides data confidentiality, semantic security, data integrity, two-party data authentication, replay protection, and weak message freshness. What's more, the protocol also has low communication overhead, for the counter state is kept at each end point and the protocol only adds 8 bytes per message. For applications requiring strong freshness, the sender creates a random nonce (an unpredictable 64-bit value) and includes it in the request message to the receiver. The receiver generates the response message and includes the nonce in the MAC computation. µTESLA provides authenticated broadcast for severely resource-constrained environments. µTESLA constructs authenticated broadcast from symmetric primitives, but introduces asymmetry with delayed key disclosure and one-way function key chains. SPINS realizes an authenticated routing application and a security two-party key agreement with SNEP and µTESLA separately with low storage, calculation and communication consumption. However, SPINS still have some underlying problems. Due to use the pairwise key pre-distribution scheme in the security routing protocol, SPINS rely on the base station excessively.

Deng et al. [4] suggested an Intrusion-tolerant routing protocol for wireless sensor networks (INSENS). This protocol leverages some concepts of SPINS to implement intrusion-tolerant multi-hop routing for WSNs. For example, we utilize keyed message authentication codes (MAC) similar to SNEP to verify the integrity of control packets. We also employ the concept of a one-way hash chain seen in µTESLA, but use the chain to provide one-way sequence numbers for loose authentication of the base station, rather than as the key release mechanism seen in SPINS. In addition, INSENS limits flooding of messages by allowing communication only between the base station and the sensor nodes, and by having sensor nodes drop duplicate messages. These techniques essentially limit the damage an intruder may cause. Together, these design choices ensure that an intruder may be able to take out a small part of the network, but cannot compromise the entire network. What's more, INSENS minimizes computation by performing all heavy-duty computations at the base station(s) and minimizing the role of sensor nodes in building routing tables, or dealing with security and intrusion-tolerance issues. Although INSENS limits the damage in the area around the invaders or the downstream area of the invaders, the damage range still will be very big when the internal attackers are around the base station.

Zhu et al proposed LEAP (Localized Encryption and Authentication Protocol) [5] protocol for sensor networks that is designed to support in-network processing, while at the same time restricting the security impact of a node compromise to the immediate network neighborhood of the compromised node. The design of the protocol is motivated by the observation that different types of messages exchanged between sensor nodes have different security requirements, and that a single keying mechanism is not suitable for meeting these different security requirements.

Hence, LEAP supports the establishment of four types of keys for each sensor node-an individual key shared with the base station, a pairwise key shared with another sensor node, a cluster key shared with multiple neighboring nodes, and a group key that is shared by all the nodes in the network. The protocol used for establishing and updating these keys is communication and energy efficient, and minimizes the involvement of the base station. LEAP also includes an efficient protocol for inter-node traffic authentication based on the use of one-way key chains. A salient feature of the protocol is that it supports source authentication without precluding in-network processing and passive participation. LEAP adopts different types of key to different network protocol message packages, and therefore can support multiple communication mode. What' more, this protocol weakens the role of base station. The main defect of the protocol is that the protocol increases overhead traffic, the failure of single point is serious, and the storage capacity is large. Besides, LEAP is inefficient in the formation and update of cluster key. What' more, the HELLO news in the network is written in plaintext rather than ciphertext, which may lead nodes to respond to invalid news and waste node resources.

The diagrams SPINS and LEAP use master keys in the key establishment. Which reduces the storage key in the memory nodes. However, resistance to attacks is low. Because the master key can be compromised at any time, set the key after deployment using this key can be compromised too. By adopting a symmetrical system, they are the most suitable and among the fastest in terms of calculation. Note that the symmetrical patterns are expensive in operations (if any) as key renewal and revocation using secret keys to exchange other secret keys. The problem is easier in asymmetrical patterns since the public key does not need to be secret.

III. THE PROPOSED MODEL

In the previous chapters it has been tried to give a general description of some of the state of the art algorithms available in the literature. It can be concluded that sensor network nodes are mostly deployed in unattended adversarial environment for example battlefield. Therefore it is extremely important for the applications of many sensor networks to have a security mechanism which provide authentication and confidentiality. The unique issue that needs to be considered in sensor network before selecting a key sharing approach is its impact on the effectiveness of in-network processing. The proposed algorithm support in-network processing and also provide security properties similar to those provided by pairwise key sharing schemes. Similar to leap+ and other protocol the proposed solution is also based on the observation that different types of messages exchanged between sensor nodes have different security requirements which leads us to the conclusion that a single keying mechanism is not suitable for meeting these different security requirements. Like leap+ the proposed algorithm also support the establishment of four different types of keys:

- **Individual key:** Shared with the base station
- **Pairwise key:** Shared with another sensor node
- **Cluster key:** With multiple neighboring nodes it is shared
- **Global key:** Shared by all nodes in the network

## 3.1 Assumption in-term of network and security

The following important assumption has been made while studying and designing the protocol

- A static sensor network where nodes are not mobile
- The base station work as a controller
- The power supplied to the base station is supplied with long-lasting power
- All nodes are equal in computational and communication capabilities
- Every node has enough space to store hundred of bytes of keying materials
- Nodes installation can be done both either through aerial scattering physical installation.
- In advance The immediate neighboring nodes of any sensor node are not known
- All the information a node holds becomes known to the attacker If it is compromised
- Attacks of the physical layer and media access control layer are not considered

The security requirements for each type of packets are different. Authentication is required for all types of packets whereas confidentiality is required for some types of packets. As it has been explained that a single keying mechanism is not appropriate for all the secure communications needed in sensor networks. So the algorithm support four types of keys for each sensor node:

## 3.2 Key Management

### 3.2.1 Individual Key

To have secure communication between the node and the base station this key is used. Every node in the network have its own individual key. Individual key is also important in the sense that it can be used to compute the message authentication code if the message is to be verified by the base station. This can be also used to send alert to the base station if there is any abnormal behavior observed. Base station can use individual key to encrypt any sensitive information such as keying material or special instructions to individual node. It is important to mention that individual key is preloaded into the network before deployment [6]. The individual key is generated as follows:

$$IK_u = fK_m(u)$$

Where $IK_u$ is the individual key, f is a pseudo-random function, $K_m$ is the master key, and u is any node for which we want to find individual key.

It is is also possible to save storage needed to keep all the individual keys the controller might only keep its master key. It compute the individual key on the fly when needed. The individual key has been find by encrypting the node id with the global key (master key) by using Advance encryption standard algorithm.

### 3.2.2 Establishing Pairwise keys

The most important step is to have pairwise key between nodes. It is very important from the security point of view as if the key is compromised its effect is localized. The pairwise key is shared between one-hop neighbors. A sensor node communicate with its immediate neighbor through pairwise key. The important assumption made to establish pair wise keys are:

- Node doesn't know its neighbor pairwise key before deployment. Pairwise key is created after deployment
- The nodes of the network are stationary nodes
- A node that is added to the network will discover most of its neighbors at the time of deployment.

The pairwise is generated by following these steps:

1- **Key Predistribution**

An initial key generated by the controller is loaded to each node. Each node then derive its master key as:

$$K_u = fK_{in}(u)$$

Where $K_u$ is the master key generated by the node u, f is a pseudo-random function, $K_{in}$ is the initial key, and u is any node for which we want to derive the master key.

2- **Neighbor Discovery**

Every node try to find its neighbor by broadcasting a HELLO message. This Hello message contains id of the node. Also a timer is started which fires after time Tmin. This node then wait for any node v which respond tho this HELLO with an ack message having the id of the node v. Ack from neighbor is authenticated using master key kv. The master key is derived as:

$$K_v = fK_{in}(v)$$

Where $K_v$ is the master key of node v, f is a pseudo-random function, kin is the initial key, and v is any node that want to find its master key.

3- **Pairwise key Establishment**

Any two nodes between the network let say node u and v compute the pairwise key as:

$$K_{uv} = fK_v(u)$$

Where kuv is the pairwise key between node u and v, f is a pseudo-random function, kv is the master key of node, and u is the node id of any node u.

4- **Key Erasure**

When the timer expires after Tmin, node u erases Kin and all the master keys of its neighbor which was computed during the neighbor discovery phase.

### 3.2.3 Establishing cluster Keys

Cluster key is established between a node and all its neighbors. Using cluster key a node encrypt broadcast message. To establish the cluster key any node u generate a random key and then encrypt this random key with the pairwise key already generated. Then the generated cluster key is transmitted to each neighbors.

### 3.2.4 Global key establishment

A key shared by the base station and every node is the global key. It is important and is used when the base station (controller) want to generate a confidential message.

### 3.3 Defense against Various Attacks

Routing control information is authenticated by the local broadcast authentication scheme which prevent most of the outsider attacks. The possible attacks an adversary may launch in the hope of creating routing loops, repelling or attracting network traffic or generating error messages are

- Spoof
- Alter
- Replay routing information
- Selective forwarding Attack
- HELLO Flood Attack
- Node cloning Attack

This scheme may not prevent the adversary from launching these attacks, however; it can thwart or minimize their consequences. The local broadcast authentication scheme makes these attacks possible only in a one-hope neighbor and hence the effect is strictly localized. This localizing effect also help in detecting these kinds of attacks. The alter attack can also be detected as the sending node may overhear the message being altered during its forwarding stage by the compromised node. It is also interesting to mention that if a node is compromised and is detected the re-keying scheme can efficiently revoke the node from the network. The hello flood attack is also possible in which an adversary may try to send a HELLO message to every node with a high transmission power so that it convince all the nodes that it is their neighbor. Here the Hello flood attack will not be succeeded beyond the neighbor discovery phase because every node accepts packets only from neighbors who are authenticated. The same is the case with node cloning attack it cannot go beyond the neighbor discovery phase.

### 3.4 Proposed model to secure the global key

The critical assumption that leap+ has considered is that within Tmin a node cannot be compromised. This idea seems practical but only in an extreme ideal condition. There is possibility that Tmin would in reality be greater than the one assumed. As an example if nodes are dropped and scattered from airplanes, the scattered nodes may arrive in different parts of the network at different times even if dropped simultaneously and hence will need some time to setup the network and exchange pairwise key. Taking the advantage of this an adversary may observe a node and obtains the key and if the global key is compromised the whole network becomes at risk. Since this is a very serious threat to security different algorithms has been studied and proposed a model that detect the compromised node and take necessary action to delete the compromised node from the network [7] [8] [9] [10] [11]. Some of the algorithms available in literature to detect compromise nodes are:

- Detecting Compromised Nodes in Wireless Sensor Network by Rick Mckenzie, Min song, Mary Mathews, sachin shetty,
- A Framework for identifying Compromised Nodes in sensor Network by Qing Zhang, Ting Yu, Peng Ning
- Malicious Node Detection in wireless sensor Networks using by Idris M.Atakli, Hongbing Hu, Yu chen
- Sensor Node Compromise Detection, The location Perspective by Hui song and liang xie

To resolve this problematic, we propose a very simple and straight forward model, where the complexity has been reduced to maximum and the global key security is ensured up-to a greater extent. In this model a sequence number has been added to every node before deployment. This sequence number or nod specific number is used to find if an adversary has taken the advantage of the key establishment time and have added its own node to get information of the network. After the pairwise keys are established the Generator or the base station (who have already the information of every node deployed) send a request to the network and ask for its sequence number, every node send its sequence number to the base station. Using this information to the previously stored information if the result is positive it stay silent. If a compromised node is detected it send a broadcast message informing all the nodes about the compromised node.

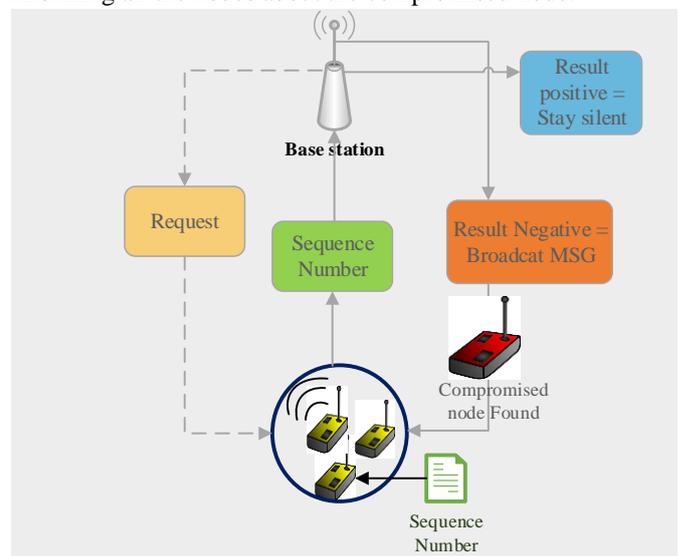

*Fig.1. Proposed Model schematic Diagram*

## IV. PERFORMANCE ANALYSIS

TOSSIM is used to simulate the Entire TinyOS applications. The goal is to achieve accurate and scalable simulation of TinyOS application. TOSSIM is a TinyOS library which work by replacing components with simulation implementations [12] [13].The key requirements for a TinyOS simulator are:

- Bridging

The simulator must work as bridge between the algorithm and implementation. It must provide developer to verify and test the code which then will run on real hardware.

- completeness

Completeness in-term of covering maximum possible system interactions and also accurately capturing behavior at a wide range of levels.

- Fidelity

The simulator must have the capability to capture the behavior of the network at a fine grain. It is important to check the timing interactions on a mote and between motes both for evaluation and testing.

- Scalability

The simulator must have the capability of handling large networks of thousands of nodes in a wide range of configurations.

### 5.1 Performance analysis

### 5.1 Pairwise key generation Time Analysis

As the time available for the generation of pairwise key is short therefore the time for the successful generation of the pairwise key with different nodes model has been analyzed. The main interest was to check if the density of network has any effect in the successful generation of pairwise key with in the available time. Interestingly it has been noticed that the algorithm successfully generate pairwise key on different node model within the available time. It has been also noticed that each pair of neighbor successfully generate the same pairwise key that is uniquely identified to the specific pair of nodes. The experiment has been performed with a difference of two, five and ten nodes repeatedly ten times each. Almost the same result has been noticed for each experimental model with successful generation of the pairwise key.

- Performed with a difference of 10 nodes ten times repeatedly

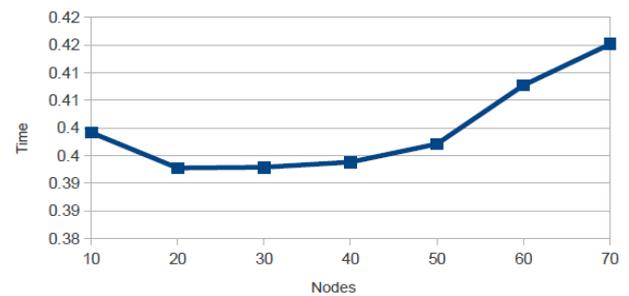

*Fig .3. Pairwise key generation time analysis*

### 5.2 Individual key generation Time analysis

The most important task was to check the successful completion of pairwise key generation. Apart from that the time for generation of individual keys for different node models has also been analyzed. It has been confirmed that each pair of node successfully generate a unique individual key.

- Performed with a sample to 20 Nodes

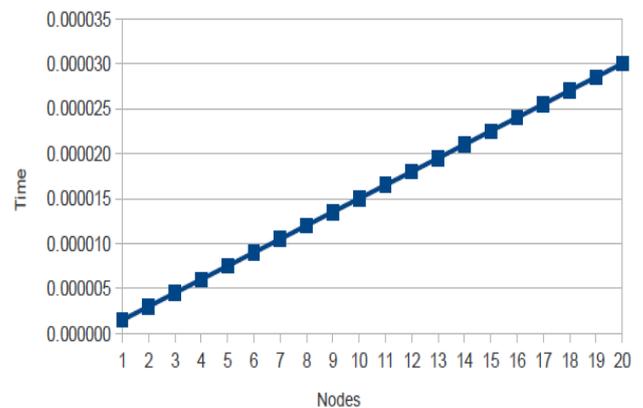

*Fig .5. Individual key generation Time analysis*

## CONCLUSION

Security is the most important issue for wireless sensor networks because of the fact of its applications. It can be setup in a very critical systems for example in hospitals, airports, military applications, burglar alarms, environment control, smart homes and traffic surveillance. The main focus of this work is to address the issue of security in WSNs. Classic security mechanism has to be avoided in WSNs due to limited constraints. Thus it has been concluded to use symmetric shared key schemes. One of the important observation for any type of key-management scheme is that a single keying mechanism is not suitable for meeting different security requirements. During this work different state of the art algorithms has been studied in depth for the proposal of a key-management scheme based on symmetric shared keys. The rapid growth of WSNs in different areas especially critical system is attracting a large number of researchers to work on its security.